# A study of the Heliocentric dependence of Shock Standoff Distance and Geometry using 2.5D MHD Simulations of CME-driven shocks

Running Heads: 2.5D Simulated CME shock-sheath distance

N. P. Savani[1,6 †], D. Shiota[2], K. Kusano[3, 7], A. Vourlidas[4], N. Lugaz[5]

1. University Corporation for Atmospheric Research (UCAR), Boulder, Co, USA
2. Computational Astrophysics Laboratory, Advanced Science Institute, RIKEN, 2-1, Hirosawa, Wako, Saitama 351-0198, Japan
3. Solar-Terrestrial Environment Laboratory, Nagoya University, Nagoya 464–8601, Japan
4. Space Science Division, Naval Research Laboratory, Washington, DC 20375-5352, USA
5. Experimental Space Plasma Group, University of New Hampshire, USA
6. NASA Goddard space flight center, Maryland, USA
7. Japan Agency for Marine-Earth Science and Technology, Yokohama, Kanagawa 236-0001, Japan

Email: neel.savani02@imperial.ac.uk

[†] This work was in part performed during a JSPS fellowship at Nagoya University

Abstract

We perform four numerical magnetohydrodynamic simulations in 2.5 dimensions (2.5D) of fast Coronal Mass Ejections (CMEs) and their associated shock fronts between 10Rs and 300Rs. We investigate the relative change in the shock standoff distance, Δ, as a fraction of the CME radial half-width, $D_{ob}$ (i.e. Δ/ $D_{ob}$). Previous hydrodynamic studies have related the shock standoff distance for Earth's magnetosphere to the density compression ratio (DR, $\rho_u/\rho_d$) measured across the bow shock (Spreiter, Summers, & Alksne 1966). The DR coefficient, $k_{dr}$, which is the proportionality constant between the relative standoff distance (Δ/ $D_{ob}$) and the compression ratio, was semi-empirically estimated as 1.1. For CMEs, we show that this value varies linearly as a function of heliocentric distance and changes significantly for different radii of curvature of the CME's leading edge. We find that a value of 0.8±0.1 is more appropriate for small heliocentric distances (<30Rs) which corresponds to the spherical geometry of a magnetosphere presented by Seiff (1962). As the CME propagates its cross section becomes more oblate and the $k_{dr}$ value increases linearly with heliocentric distance, such that $k_{dr}$= 1.1 is most appropriate at a heliocentric distance of about 80Rs. For terrestrial distances (215Rs) we estimate $k_{dr}$= 1.8 ±0.3, which also indicates that the CME cross-sectional structure is generally more oblate than that of Earth's magnetosphere. These alterations to the proportionality coefficients may serve to improve investigations into the estimates of the magnetic field in the corona upstream of a CME as well as the aspect ratio of CMEs as measured in situ.

# 1. Introduction

Although the properties of CMEs have been studied for many years (e.g. Gopalswamy et al. 2009; St Cyr et al. 2000; Vourlidas et al. 2010; Yashiro et al. 2004), it is only recently with the dawn of the STEREO (Eyles et al. 2009; Kaiser et al. 2008) and Coriolis (Eyles et al. 2003; Jackson et al. 2004) missions that their white light structures have been studied much deeper into interplanetary space between the Sun and the Earth (Maloney & Gallagher 2011; Savani et al. 2009). Even then, the white light structures become too dim within the cameras to accurately trace out the global shape as they approach Earth. As such, novel methods to estimate the cross section at terrestrial distances have been investigated. Studies to estimate the cross sectional morphology of CMEs using one dimensional (1D) in situ data have recently been investigated (Russell & Mulligan 2002; Savani et al. 2011b). However these studies not only rely on the properties of the shock front associated to the CME but also assume the geometry and shock standoff distance is identical to Earth's magnetosphere. By using physical arguments, a study by Siscoe and Odstrcil (2008) suggest that this may not be the case. In this paper we test the validity of assuming a magnetosheath as an appropriate geometry for a sheath region ahead of a CME.

CMEs observed remotely from coronagraphs after they are initially launched off the Sun are often treated as objects with circular cross-sections (Howard et al. 1982; Krall et al. 2001; Vourlidas, et al. 2010). This circular shape is consistent with theoretical derivations of twisted magnetic field lines that form flux rope structures (Chen 1989; Chen et al. 1997; Titov & Demoulin 1999), which are also used to initiate CMEs in computational models (Lugaz, Manchester, & Gombosi 2005; Manchester et al. 2004; Shiota et al. 2005; Shiota et al. 2010). These circular structures (i.e. aspect ratio of 1) found in force-free conditions are often considered to be a correct 'zeroth' order approximation for a class of ICME known as magnetic clouds, which are observed in situ to have a smooth field line rotation consistent with a flux rope. Some of the earliest modelling of in situ data assumed circular cross sections (Burlaga 1988; Lepping, Jones, & Burlaga 1990), and this geometry currently remains as the zeroth order approximation for estimates of their total magnetic flux content at terrestrial distances when considering long term space climate variability and potential space weather effects (Lockwood et al. 2004; Owens & Crooker 2006). However, as a CME propagates at a near constant radial velocity while expanding in the meridional direction to maintain a near-constant angular width, it has long been suspected that a CME could flatten into an elliptical cross-section as it travels further into the heliosphere (Riley & Crooker 2004; Savani et al. 2011a). As such, numerous attempts have been made to modify the static picture of a simple constant-$\alpha$ flux rope fitting (Hidalgo et al. 2002; Marubashi 1986; Mulligan & Russell 2001; Owens, Merkin, & Riley 2006).

Russell and Mulligan (2002) suggested that the aspect ratio of a CME may be estimated from measuring the distance between the shock and the leading edge of the CME (shock standoff distance, Δ) and then by assuming that the geometry of Earth's bow shock is appropriate for interplanetary CMEs (ICMEs; see figure 1). The authors therefore began the derivation of their estimates from the semi-empirical formula (Spreiter, et al. 1966),

$$\frac{\Delta}{D_{OB}} = 1.1\ \frac{\rho_u}{\rho_d} \qquad (1)$$

which is based on gas-dynamic theory for Earth's magnetosphere, where $D_{OB}$ is the distance from the center to the nose of the obstacle (originally the Earth's magnetopause), ρ is the plasma density and the subscripts 'u' and 'd' denote up and downstream of the shock, respectively. $D_{OB}$ was then later substituted for the vertical extent of the obstacle ($D_T$, see section 2 for details). The density compression ratio (DR) coefficient, $k_{dr}$ =1.1 was calculated experimentally under laboratory conditions. A metallic object, oblate in shape, was fired through neutral argon gas and snapshots in time were taken during its propagation. This allowed measurements to be taken for the shape and location of the shock front in relation to the obstacle under different firing speeds, which correlate to the compression of gas across the shock front. As much of the theory that correlates this sheath geometry to a CME is fundamentally based on this empirical equation in this paper (see section 2 for details), we investigate the robustness of $k_{dr}$ for fast CMEs travelling between the Sun and 1AU.

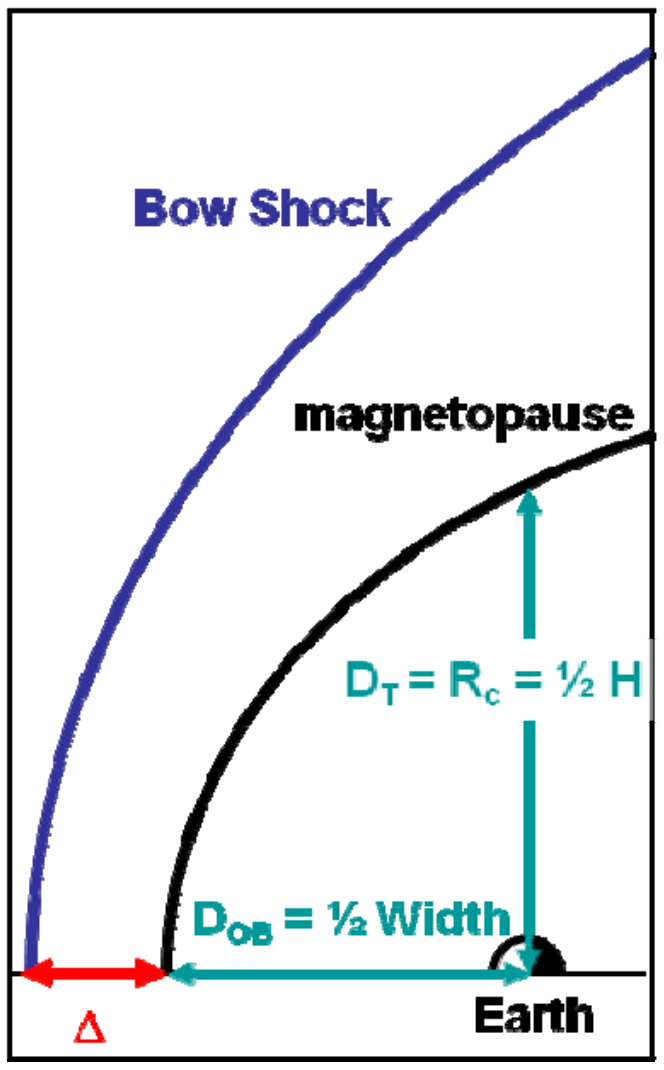

Figure 1. Schematic displaying the relationship between Earth's bow shock and characteristics of an ICME. We relate the width of a CME measured with in situ instruments and the distance from the Earth to the magnetopause nose ($D_{OB}$), and the vertical size (1/2 H) to the radius of curvature ($R_c$). (Taken from Savani, et al. 2011b).

Russell and Mulligan (2002) investigated a CME event from Pioneer Venus Obiter (PVO) during the spacecraft's transit to Venus. The authors used in situ data to measure the shock standoff distance, Δ, and related it to equation 1 in order to predict the aspect ratio of the CME's cross section. The authors estimated an aspect ratio of 4, which is in agreement to the geometrical prediction by Savani, et al. (2011a). However, a more thorough investigation of this mathematical formulism by Savani, et al. (2011b) for 45 events between the ~0.5 and 5.5 AU found an average aspect ratio of 2.8 ± 0.5. From these results, Savani, et al. (2011b) concluded that deformations to the leading edge of CMEs may occur frequently, thereby predicting a smaller radius of curvature. But the authors noted that the estimates are predicted on the assumption that the behaviour of the sheath is identical for interplanetary CMEs and Earth's magnetosheath. This disparity was highlighted on physical grounds by Siscoe and Odstrcil (2008). They argued that the transient nature of CMEs and their lateral expansion changes the sheath behaviour by prohibiting the solar wind from deflecting around the CME. The authors suggest that the solar wind plasma 'piles up' ahead of the CME, thereby categorising this as an 'expansion sheath' apposed to the 'propagation sheath' for common magnetospheres. Therefore these authors would predict that the coefficent, $k_{dr}$ would not necessarily equal 1.1 and it may perhaps vary as the CME becomes more elliptical (i.e. for larger heliocentric distances).

Analysis based on similar theoretical geometry of equation 1 and detailed in section 2 have been investigated beyond the heliocentric range of Savani, et al. (2011b). Lynnyk et al. (2011) investigated the radius of curvature of CMEs in a similar manner to Russell and Mulligan (2002) and Savani, et al. (2011b), however they concentrated on CMEs detected by the Voyager mission (Burlaga et al. 1981) at much larger heliocentric distances (1 – 30AU).

Also, the CME geometry based on equation 1, formed the foundation for deducing the magnetic field strength upstream of the CME-driven shock in the solar corona between 1.5 – 23 solar radii (Gopalswamy & Yashiro 2011). In this scenario, the shape of the CME and sheath distance was measured remotely off the limb of the Sun, allowing an estimate of the sound speed Mach number to be made. With further assumptions, an estimate of the magnetic field strength was made.

Therefore, it is for the reason of multiple research interests, all based on a single semi-empirical formula (equation 1), which has itself been considered questionable, that we investigate the suitability of $k_{dr} = 1.1$. We carry out our investigation with the aid of 2.5D MHD simulations detailed in section 3.

# 2. Theory

Spreiter et al. (1966) measured the densities up and downstream as well as the position of the bow shock from an experimental procedure designed to mimic the terrestrial magnetosphere with the aid of an oblate metallic obstacle fired through argon gas. They concluded that $k_{dr} = 1.1$ remains robust for a variety of upstream solar wind velocities (Mach number, M, between 5 and 100) and the ratio of specific heats (polytropic index), $\gamma$ between 1.1 and 2.

However, the shape of the obstacle was shown to make a significant impact. Seiff (1962) performed a similar analysis for a spherical obstacle and produced the relation,

$$\frac{\Delta}{D_{OB}} = 0.78 \frac{\rho_u}{\rho_d} \quad . \tag{2}$$

The practical application of equations 1 and 2 is carried out by exchanging the ratio of densities to Mach number. This can be done under hydrodynamic (HD) conditions by invoking the conservation of mass, to produce the nontrivial solution (Priest 1984)

$$\frac{\rho_u}{\rho_d} = \frac{(\gamma-1)M_u^2+2}{(\gamma+1)M_u^2} \quad , \tag{3}$$

in terms of the sonic Mach number $M_u \equiv v_u/c_{s,u}$ , where $c_{s,u} \equiv (\gamma p_u/\rho_u)^{1/2}$. Here, p is the pressure and v is the velocity. However, under the more sophisticated magneto-hydrodynamic (MHD) regime, equation 3 is only applicable for a parallel shock where the velocity and magnetic field are both parallel. While the MHD solution for a perpendicular shock where the velocity is normal to the shock front follows (Priest & Forbes 2007):

$$2(2-\gamma)\ \mathbf{X}^2 + [2\gamma\beta + (\gamma-1)\gamma\beta\, M_u^2 + 2\gamma]\ \mathbf{X} - \gamma(\gamma+1)\beta\, M_u^2 = 0\ , \tag{4}$$

where $\mathbf{X} = \rho_d/\rho_u$ and β is the upstream plasma beta. Equation 4 is a quadratic equation that can be solved as a function of Mach number, with the magnetic field serving to reduce **X** below its hydrodynamic value. However for the purposes of CME-driven shocks, current studies have focused on the application of only the HD case from equation 3. Unfortunately under a low Mach number regime ($M \lesssim 3$) neither equation 1, 2 nor 3 are valid. This led Farris and Russell (1994) to make an adjustment to the denominator of equation 3 from $(\gamma+1)M_u^2 \rightarrow (\gamma+1)(M_u^2-1)$. This was done on an intuitive basis.

Even though the Mach number is the observationally measured estimate used to infer the CME-driven shock properties, we choose to focus on quantifying $k_{dr}$ in the work presented here because the inferred answers assume the less accurate HD scenario.

Farris and Russell (1994) noted that the radius of curvature at the nose of Earth's magnetopause can be estimated as equal to the vertical extent of a magnetosphere (sometimes called the terminator distance). Under this assumption the distance from the center of the Earth to the nose of the magnetopause can be used to estimate the radius of curvature:

$$\frac{R_C}{D_{OB}} = 1.35\ . \tag{5}$$

While, in fact the radius of curvature stated by Spreiter, et al. (1966) is, $R_c = D_{OB}\left(3 + \sqrt{21}\right)/6$, and is approximately 1.26 (Verigin et al. 2003).

By combining equations 1, 3 and 5 (i.e. under the HD regime) we are able to relate the radius of curvature of an obstacle based on an oblate magnetosphere to two measurable quantities,

$$\frac{\Delta}{R_c} = 0.81\ \frac{(\gamma-1)M_u^2+2}{(\gamma+1)M_u^2}\quad ,$$

$$for\ \ \gamma = {}^5\!/_3 \ \Rightarrow\ \frac{\Delta}{R_c} = 0.204 + 0.611\ M_u^{-2}\ . \qquad (6)$$

However, the radius of curvature may be related to the spherical geometry found in equation 2 to produce the formulae used by Russell and Mulligan (2002):

$$\frac{\Delta}{R_c} = 0.58\ \frac{(\gamma-1)M_u^2+2}{(\gamma+1)M_u^2}\quad ,$$

$$for\ \ \gamma = {}^5\!/_3 \ \Rightarrow\ \frac{\Delta}{R_c} = 0.195 + 0.585\ M_u^{-2}\ . \qquad (7)$$

The authors performed the calculations with an assumed polytropic index of 5/3. The spherical geometry inherent in equation 7 is perhaps more applicable than equation 6 for CMEs early in their propagation phase, when the CME morphology may be considered to have a more idealised circular cross-section. At a heliocentric distance of ~1.4Rs, Gopalswamy et al. (2012) showed that the coronal magnetic field can be indirectly estimated by fitting a circular shape to a CME observed in SDO and by implementing equation 6. However this analysis required a further assumption that the Alfven Mach number ($M_A$) can be substituted for the gasdynamic sonic Mach number (M), which is a good approximation only for low Mach numbers in the solar wind (Fairfield et al. 2001).

# 3. Numerical Model

## 3.1 Numerical Scheme

The axisymmetric simulation performed in this work solves the MHD equations using a finite volume method with the Harten-Lax-van Leer Discontinuities (HLLD) nonlinear Riemann solver (Miyoshi & Kusano 2005), the third-order TVD Monotone Upstream-centered Schemes for Conservation Laws (MUSCL), and the second order Runge-Kutta time integration. Details of the numerical model are specified in Shiota et al. (2008). The MHD equations implemented in the system follow the equation of continuity, the equation of motion, the induction equation, and equation of energy with gravity.

However, this simulation alone without any corrections may violate the divergence-free condition of the magnetic field. As the magnetic field is derived from the solenoidal condition, the simulation may breakdown if the numerical value of $\nabla.\boldsymbol{B}$ is not sufficiently small. This is corrected in our simulation by employing the hyperbolic divergence B diffusion method (Dedner et al. 2002) and detailed in Shiota, et al. (2008). In essence, once a finite value of $\nabla.\boldsymbol{B}$ is locally formed within the numerical domain, the quantity is distributed across the full domain and thereby reduces any localised magnitude.

The numerical domain is axisymmetric in the meridional plane and extends in the radial direction between 4 Rs and 304 Rs. The domain is discretized to a grid of (nr, nθ)=(1604,512) in a non-uniform manner in order to enhance the resolution at smaller heliocentric distances (Shiota, et al. 2008). The inner boundary is set outside the sonic point at 4 Rs. The thermodynamic effects such as initial coronal heating, radiative cooling and initial acceleration processes for the solar wind are neglected in this study. The electric resistivity is set to be zero, thus causing magnetic reconnection to occur only as a result of numerical diffusion.

## 3.2 Solar wind

To obtain the initial steady-state solar wind solutions, we begin with reasonable initial conditions and integrate the MHD equations (Shiota, et al. 2008) in a Cartesian coordinate system while the grid system remains in a spherical system. Initially, a supersonic solar wind is set to be at a uniform speed of 300km/s, with a uniform density of 6.33x $10^3$cm$^{-3}$ and at a uniform temperature of 1.32x $10^6$K. The initial magnetic field of the Sun is considered to have a simple dipolar configuration. However, as the inner boundary can be considered to be past the source surface (which is usually set at ~2.5Rs), we assume all the field lines are 'open' and directed radially from the Sun. The field direction is flipped at the equator and follows

$$\boldsymbol{B}(X,Z) = B_r(X,Z)\,\boldsymbol{e}_r = \frac{B_{r0}}{r^2}\ \tanh\left(\frac{1}{w_{cs}}\frac{Z}{r}\right)\ . \quad (8)$$

Where $r = \sqrt{X^2 + Z^2}$, $B_{r0}$ = 3 G, and the current sheet thickness is defined by $w_{cs}$= 0.05. Therefore a field strength of 0.1875 Gauss is injected at the inner boundaries of the poles.

The specific heat ratio, γ is set to 1.4 in order to simply introduce a small energy source. The heat ratio is set slightly below the average estimate for the solar wind (γ=1.46) as measured in situ (Kartalev et al.

2006; Totten, Freeman, & Arya 1995). As the work presented below investigates the DR coefficient, $k_{dr}$ as defined in equation 1, and the maximum theoretical density compression ratio is intimately linked to $\gamma$ (i.e. a maximum ratio of 4 when $\gamma$ =5/3), it is important for this study to estimate the compressibility of the solar wind appropriately. Many MHD simulations use a reduced non-adiabatic polytropic index of $\gamma$=1.05 as a simplified correction for neglecting the thermodynamic effects in the corona (e.g. Linker et al. 1999), however this is not appropriate for our study, therefore we must start our simulation domain above these effects (Pomoell & Vainio 2012). For this reason we set the inner boundary of our domain at 4Rs and fix the initial CME flux rope at 10Rs (see section 3.3 for details on flux rope initiation). It is worth noting that Pomoell, Vainio, and kissmann (2011) studied the dynamics of shocks by setting the heating source term to an appropriate value, such that the polytropic index may be maintain at 5/3 while retaining an identical background solar wind solution produced by a polytropic index of 1.05.

We also note that 2.5D simulations have been shown to provide comparable results to their 3D equivalent if their initial momentum are identical (Jacobs, van der Holst, & Poedts 2007). Jacobs, et al. (2007) showed that the position of the front and center of mass were comparable to their 3D equivalent. They predicted that the 2.5D simulation would be subject to a smaller drag force because the CME surface is spread over a larger area when compared to the 3D case. Because the domain of the authors work was limited to a maximum heliocentric distance of 30Rs, they predicted the consequence of the smaller drag force would result in an earlier arrival time of the CME at Earth. However their simulation is based on a 'density-driven' model which involves the launching of a high-density plasma blob, and therefore does not consider the magnetic forces.

Our solar wind model has similar parameters to the study performed by Tsurutani et al. (2003); these authors investigated the propagation of a CME-driven shock. Both models are axisymmetric and neither include a bimodal wind nor a spiral interplanetary magnetic field. However Tsurutani, et al. (2003) investigated a spatial regime starting at the solar surface, as such they chose to vary the polytropic index with radial distance in order to accommodate the additional heating source processes occurring in the solar corona. Manchester et al. (2005) compared their findings from a global 3D model with a Gibson-Low magnetic flux rope with that of Tsurutani and co-authors, and found a similar pattern for the shock normal direction at low latitudes. The shock Mach number is found to increase to about 18Rs, whereas Tsurutani, et al. (2003) find that the Mach number aproaches a maximum at 130Rs.

### 3.3 CME creation

As we are interested in the morphology of the CME and the shock position, we do not consider the trigger process or possible instability conditions for the launch of CMEs (e.g. Antiochos, DeVore, & Klimchuk 1999; Chen & Shibata 2000; Forbes & Priest 1995; Inoue & Kusano 2006; Kusano et al. 2004; Roussev, Lugaz, & Sokolov 2007). We artificially superimpose a toroidal flux rope onto the background solar wind as a non-equilibirum structure. The magnetic field inside the flux rope is defined as

$$\boldsymbol{B} = \alpha\nabla \times (\Phi e_{\emptyset}) + \nabla \times \nabla \times (\Phi e_{\emptyset}) \quad , \tag{9}$$

Where $e_\phi$ is the unit vector in the $\phi$ direction, $\alpha$ is the force free parameter and $\Phi$ is the flux function. The magnetic field is semi-analytically calculated by the axisymmetric Grad-Shafranov equation,

$$\frac{\partial^2\psi}{\partial X^2} + \frac{\partial^2\psi}{\partial Z^2} + \left(\frac{1}{4X^2} + \alpha^2\right)\psi = 0 \tag{10}$$

where $\psi = \sqrt{X}\Phi$ (similar to the toroidal solution by Miller & Turner 1981) and by applying a boundary condition of

$$\Phi = 0 \quad \text{when} \quad r = \sqrt{(X - X_c)^2 + Z^2} \geq a_c \quad . \tag{11}$$

The variables $X_c$ is the initial heliocentric distance of the flux rope center; $a_c$ is the minor radius of the flux rope; and $\alpha = 1/X_c$. The Grad-Shafranov equation is solved numerically using the successive over-relaxation (SOR) method (Press, et al. 1992) by following,

$$\psi_{i,j}^{n+1} = \psi_{i,j}^{n} + \omega\left[\frac{\psi_{i+1,j}^{n}+\psi_{i-1,j}^{n}}{D\,\Delta X^2} + \frac{\psi_{i,j+1}^{n}+\psi_{i,j-1}^{n}}{D\,\Delta Z^2} - \psi_{i,j}^{n}\right] \tag{12}$$
$$D = \left[\frac{1}{4X^2} - \frac{2}{\Delta X^2} - \frac{2}{\Delta Z^2} + \alpha^2\right].$$

Where $\omega$, set to equal 1.5, is the extrapolation factor chosen to accelerate the rate of convergence of the iterates to the solution. The initial plasma density of the flux rope is also defined as twice the surrounding solar wind value. The magnetic flux threading through the cross-section, $F_L = \iint \boldsymbol{B}_\phi \; dX\, dZ$. Everywhere within the flux rope is set to an approximately force-free condition in the initial state. The values of the parameters investigated in the paper are displayed in table 1.

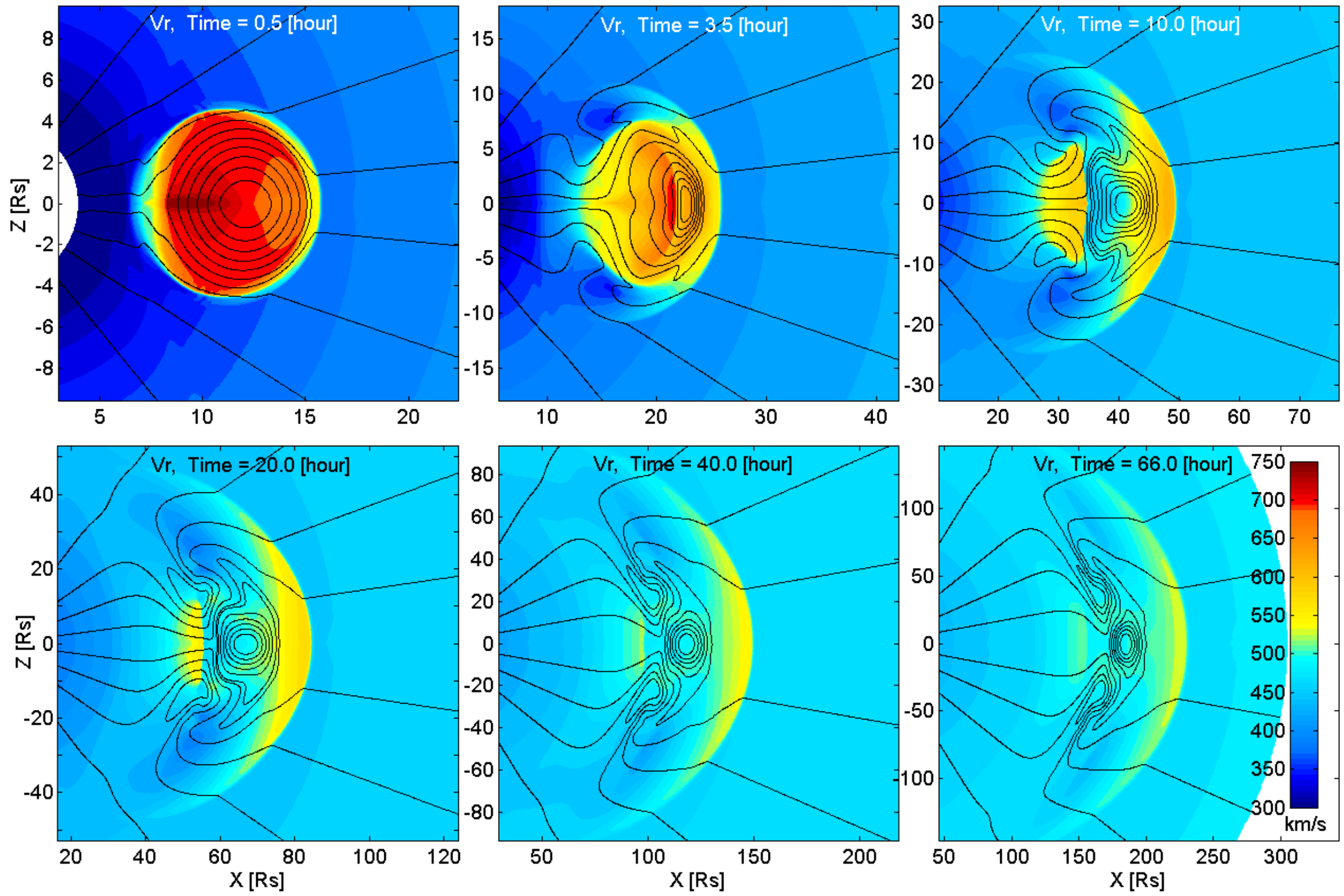

Figure 2. 2.5D simulated flux rope for case study 1 is displayed at six frames during its propagation from 10Rs to ~200Rs. The radial component of the velocity is shown in color within the simulated domain (between 4Rs and 304Rs). The color bar is the same for all the panels of each of the parameters. The black curves follow the contours of constant flux function, which trace out the magnetic field lines. Note that the magnetic field lines in each panel are different to each other.

# 4. Results

In this paper we use four simulated case-study events to investigate the impact of varying the initial speed and the size of the flux rope. Case 1 is our benchmark event that begins at 10Rs and has a radial width of 4Rs; see figure 2. This size was chosen by the restrictions imposed by the assumptions used in the simulation and observational evidence. A restriction to the minimum heliocentric distance for the FR is set from the inability to correctly consider the effects of coronal heating and the angular width of approximately 50 degrees is estimated as a typical width of a CME from remote sensing observations (St Cyr, et al. 2000). The magnetic flux contained within the FR was set to 3x $10^{21}$Mx which is of a typical order of magnitude for observed in situ measurements (Kataoka et al. 2009). The magnetic field magnitudes at Earth for these events are ~90nT which would correspond to very large geo-effective events (Manchester et al. 2006). The FR is launched with a uniform radial velocity of 700km/s. The other events vary these parameters as shown in table 1. Because our simulation is limited to introducing a FR with an instantaneous speed rather than with a more gradual acceleration, we are limited to a maximum initial speed of about 1200km/s (Case 4). Any further increase in speed is found to build up an excess in magnetic pressure early in the propagation of the FR such that it equilibrates by generating an unusually large radial expansion. Although this limitation prevents our current study from investigating extreme events such as the Carrington storm, it does allow us to investigate the more frequently observed fast CMEs.

| | Initial Heliocentric, $X_c$ distance, Rs | FR radial Width, $a_c$ Rs | Initial Velocity km/s | Magnetic Flux, $F_L$ Mx |
|---|---|---|---|---|
| Case 1 | 10 | 4 | 700 | $3x10^{21}$ |
| Case 2 | 10 | 2 | 700 | $3x10^{21}$ |
| Case 3 | 10 | 4 | 900 | $3x10^{21}$ |
| Case 4 | 10 | 4 | 1200 | $3x10^{21}$ |

Table 1. Initial conditions for the four 2.5D simulated CME case study events.

Figure 3 shows an exploded view of the CME from case 1. Figure 3a displays an extra contour following the flux function as a white line. This line is manually selected between the open and closed field lines from the automated contours levels shown in black. The distance between two adjacent open/closed field lines is then used to estimate the error in the vertical extent ($\varepsilon_h$) of the FR. This process was carried out for each frame that was investigated in further detail. Figure 3b displays the velocity vectors of the solar wind relative to the center of the flux rope. The theta direction of the spherical coordinate system used in these simulations is the co-latitude values (i.e. theta increases from the ecliptic north towards the ecliptic plane). The plasma within the sheath is deflected and slowed down by the leading shock front. Figure 3a and 3b clearly show the shock boundary, while 3b displays the solar wind smoothly deflecting around the FR obstacle. Figure 3c and 3d displays the phi and theta components of the current density, J. The change in the current density emphasizes the approximate location of the FR leading edge.

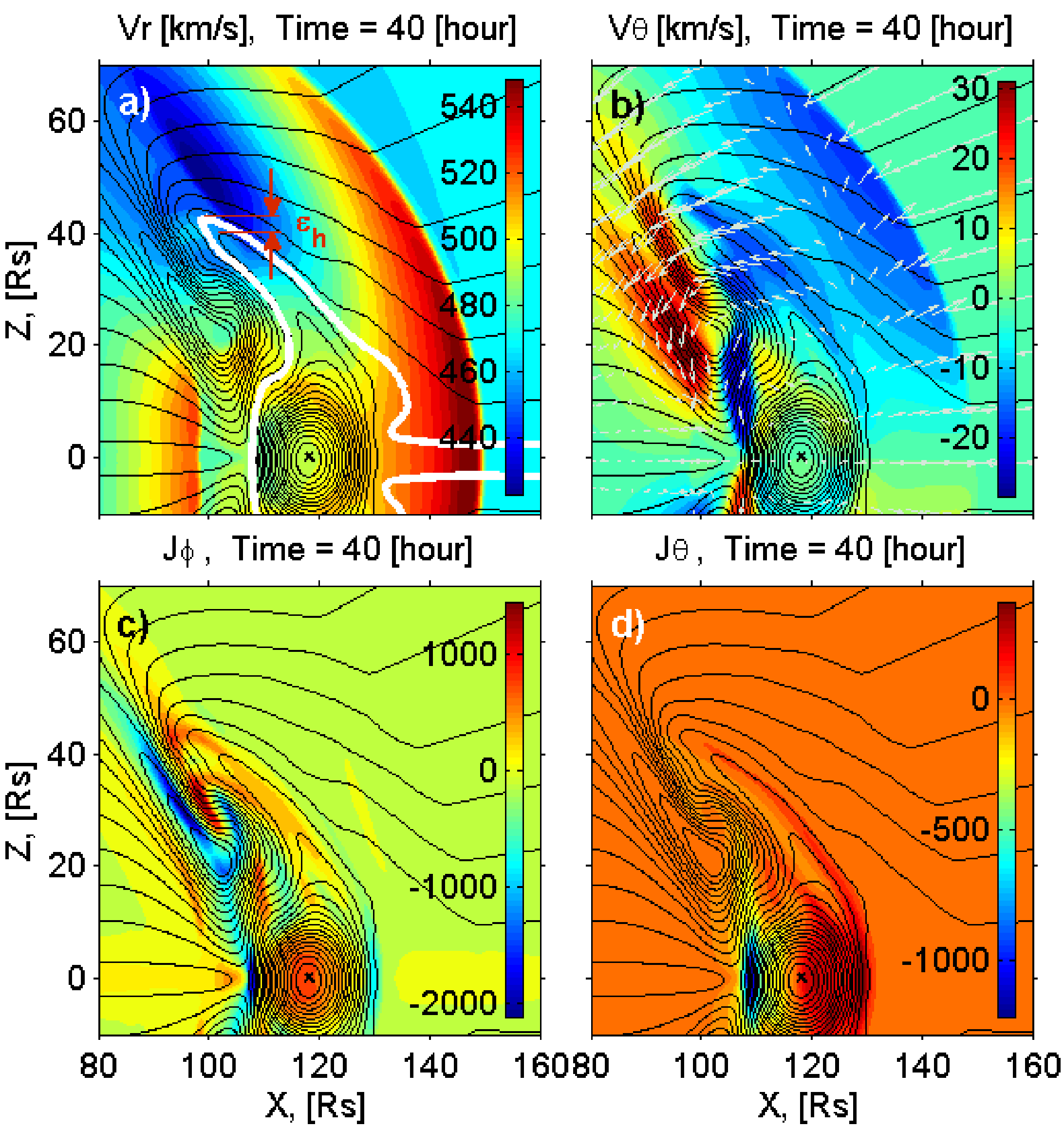

Figure 3. Displays a 2.5D simulated flux rope for case study 1 at a single time. The variables of the velocity in the radial and theta direction are displayed in panel a and b, respectively. The current densities in the phi and theta directions are shown in panels c and d, respectively. The black contours represent lines of constant flux function, Ap which trace out the magnetic field lines. The white curve in panel a displays an additional line of constant Ap which is manually selected and used to estimate the error in the flux rope height as determined by a boundary of closed/open magnetic field lines. Panel b displays vector arrows in white which display the flow direction of the solar wind referenced from the center of the flux rope.

The plasma beta within the FR for the simulations is below one; the solar wind has a beta value of approximately one and a higher value is present around the location of the current sheet. These results can be seen in figure 4. Figure 4 shows three frames during the early propagation phase of the simulated CME from case 1. The first frame displays a higher beta plasma surrounding the FR leading edge with the leading shock front having recently been formed. The higher beta value within the sheath is indicated by the red coloration and the plasma smoothly flows around the CME leading edge (green vectors). The higher beta emphasises the relative importance of the plasma forces above the magnetic forces. The sheath plasma flows over the CME to the rear side where it interacts with plasma that is more dominated by the magnetic forces (beta less than one). The latter two frames show the high beta plasma flowing behind the FR and thus 'kinking' the field lines in a manner that forms a flux tube from numerical diffusion.

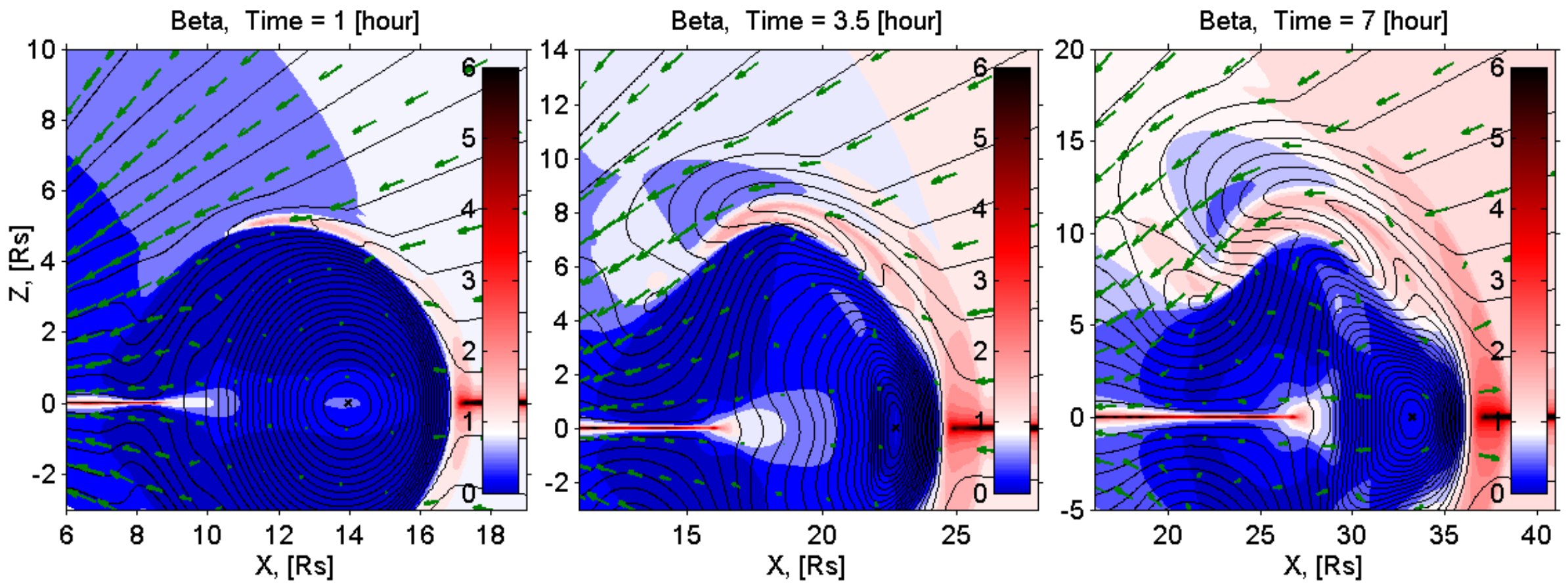

Figure 4. Displays the plasma beta for the case 1 simulated FR at 3 time frames after its injection into the solar wind: 1 hour, 3.5 hours and 7 hours. The black lines represent the magnetic field lines and the green vectors show the plasma flow vectors in relation to the motion of the FR center.

As the FR propagates, the shock front spans over a larger position angle, thereby disturbing the solar wind further north and south than the flux rope itself. However, the location of the shock front and the time evolution of the shock standoff distance are better viewed as a height-time map, shown in figure 5. The height-time maps have often been used for locating the heliocentric position of a CME from remote observations (Sheeley et al. 1999), and have more recently been used to determine their propagation direction (e.g. Davies et al. 2009; Rouillard et al. 2009). Figure 5 displays the magnetic field contours of the flux rope as black lines, while the color scale displays the radial velocity. The shock front is clearly located ahead of the FR and shown as a distinct boundary on the color scale. Behind the rear edge of the FR shows a region where the solar wind velocity changes rapidly.

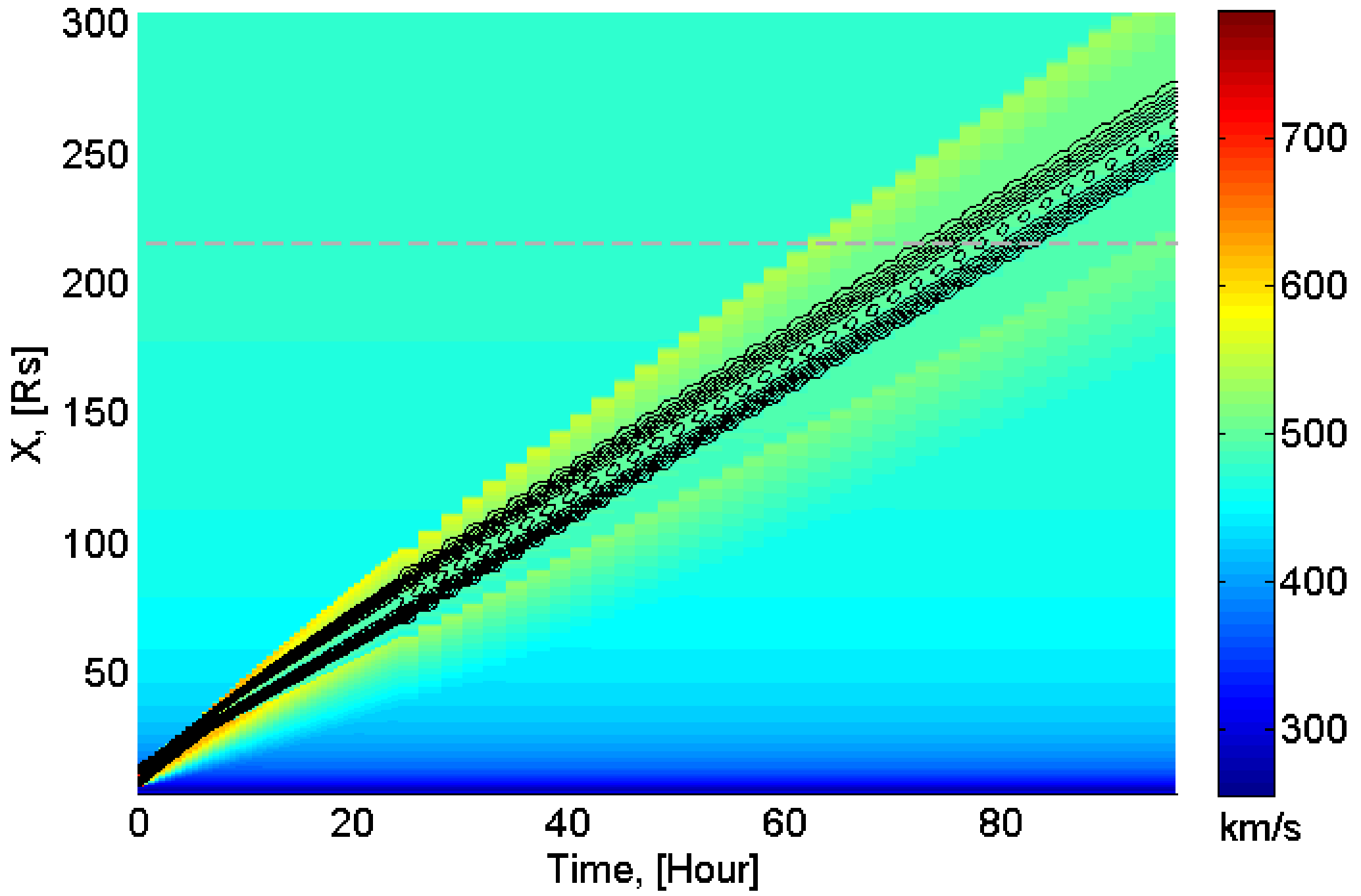

Figure 5. Height-Time map of the flux rope along the ecliptic plane. The color scale displays the radial component of the velocity and the black contours are the same field lines of the flux rope as displayed in figure 2. The grey dashed line is at a constant heliocentric of 215 Rs, indicating the location of Earth.

With the aid of Figure 6, we show that this is the reverse shock pressing onto the solar wind from rear edge of the CME. This is an effect from introducing a FR with an instantaneous fast speed. The effect can also be seen in figure 2b, between 3.5 and 10 hours, where a forward-reverse wave pair is initially formed ahead of the FR; the reverse wave then propagates through the FR resulting in a difference in speed between the front and rear edge which effectively creates a large expansion velocity for the FR. In the reference frame of the FR, this causes the rear edge of the FR to ram into the background solar wind causing a forward- reverse pair of shocks. This type forward-reverse pair behaviour has also been noted as an 'over-expansion' of a CME (Gosling et al. 1995) and occurred in uniform solar wind conditions such as our simulation; however the observations occurred at high latitudes within fast solar wind conditions. Therefore, although the global properties after ~0.5AU appear similar to those observed by Gosling et al. (1995), we believe the reasons for the forward-reverse pair are different in our simulation. This behaviour of a shock wave travelling through the CME at an early stage of its propagation (i.e. ≲30Rs) when the CME has a mostly circular cross-section may account for the slightly different magnetic structure to that simulated by Riley & Crooker (2004). In our simulations the shape of the flux rope is slightly unusual in that its lateral expansion apppears to only invovle the outer most flux (shown by the single most extended field line) which expands to maintain the nearly constant angular width (50°) of the flux rope. This small amount of flux forms a thin ‘wing’ while the majority of the flux remains nearly circular in shape rather than being evenly distorted into an oblate shape. These wings have a higher plasma beta value (greater than 1) than the central sections of the FR – see figure 4. Further investigation into this behaviour would require more detailed studies in varying the intial speeds and by incorporating a more realistic acceleration process during the injection of the CME. These extended wings also appear to trigger strong inflows behind the flux rope. This can be clearly seen in figure 3b and figure 4 with the vector arrows of the flow direction. The previously open field lines of the solar wind are drawn into a deep u-shape behind the rear edge at two locations adjacent to the wings and reconnect to form closed flux systems.

The location of the FR is shown between two vertical grey lines in the right hand panel of Figure 6. Figure 6 displays the result of a 1D spatial cut through the ecliptic. Therefore in this figure, we display the forward shock to the right (i.e. at a larger heliocentric distance) of the reverse shock (indicated by large discontinuities in the velocity profile), whereas it is often observed the other way for a time series of in situ measurements. We choose to display our figure this way because we do not focus on the specific nature of the forward-reverse pair, but instead are interested in the spatial locations of the FR and its associated forward shock front ahead.

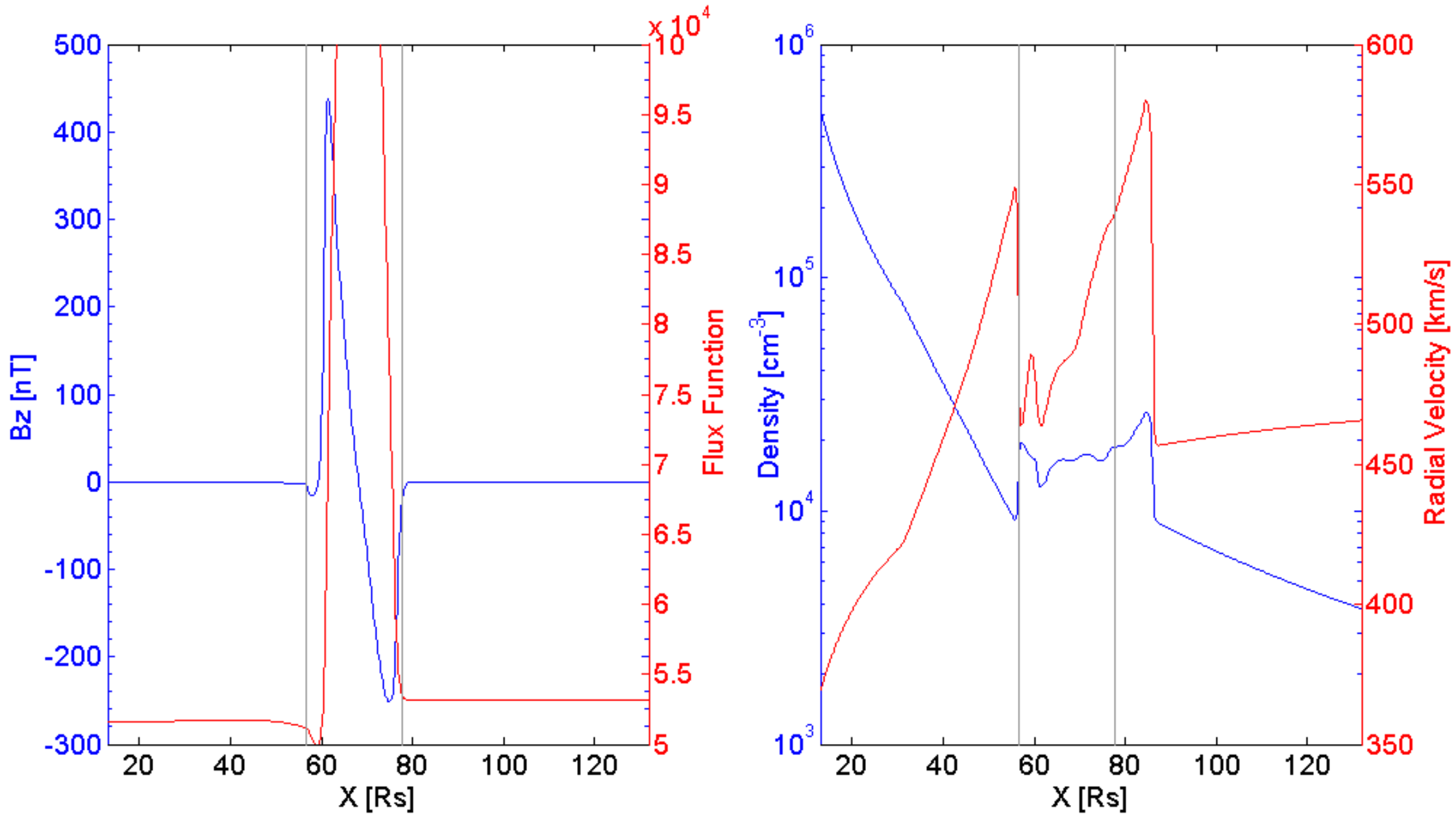

Figure 6. A 1D cut of the simulated domain from Case 1, when the FR is located at approximately 70Rs. The spatial profile was taken along the ecliptic. The left panel displays the magnetic field component (Bz, blue) that is perpendicular to the ecliptic plane and the flux function (red). Right panel: shows the density of the solar wind plasma (blue) and the radial component of the velocity (red).

The heliocentric position of the shock and the flux rope boundary were manually selected within fixed frames. For the four simulations, between 15 and 20 frames were selected, and the flux rope boundary was defined by four points: two along the ecliptic which measured the radial width of the flux rope by inspecting the flux function in figure 6; and two perpendicular to the radial which measures the vertical height (Chen, et al. 1997; Nieves-Chinchilla et al. 2012; Savani et al. 2012), which were estimated from frames similar to figure 2. Figure 7 tracks the increase in the flux rope's radial width, the perpendicular vertical height and the radial width of the sheath (the shock standoff distance). The vertical height and the sheath width can be clearly seen to increase in size at a faster rate than the radial width of the flux rope; however this is more prominent with the vertical size than the sheath distance. The small increase in the sheath distance relative to the radial width was also observed by Manchester, et al. (2005), and can be qualitatively explained by two potential factors: 1. As the flux rope propagates, it decelerates and thereby decreases the shock strength. So, as the flux rope moves through the heliosphere, the Mach number decreases and the shock standoff distance increases. 2. The morphology of the flux rope changes therefore the dynamics of the standoff distance changes.

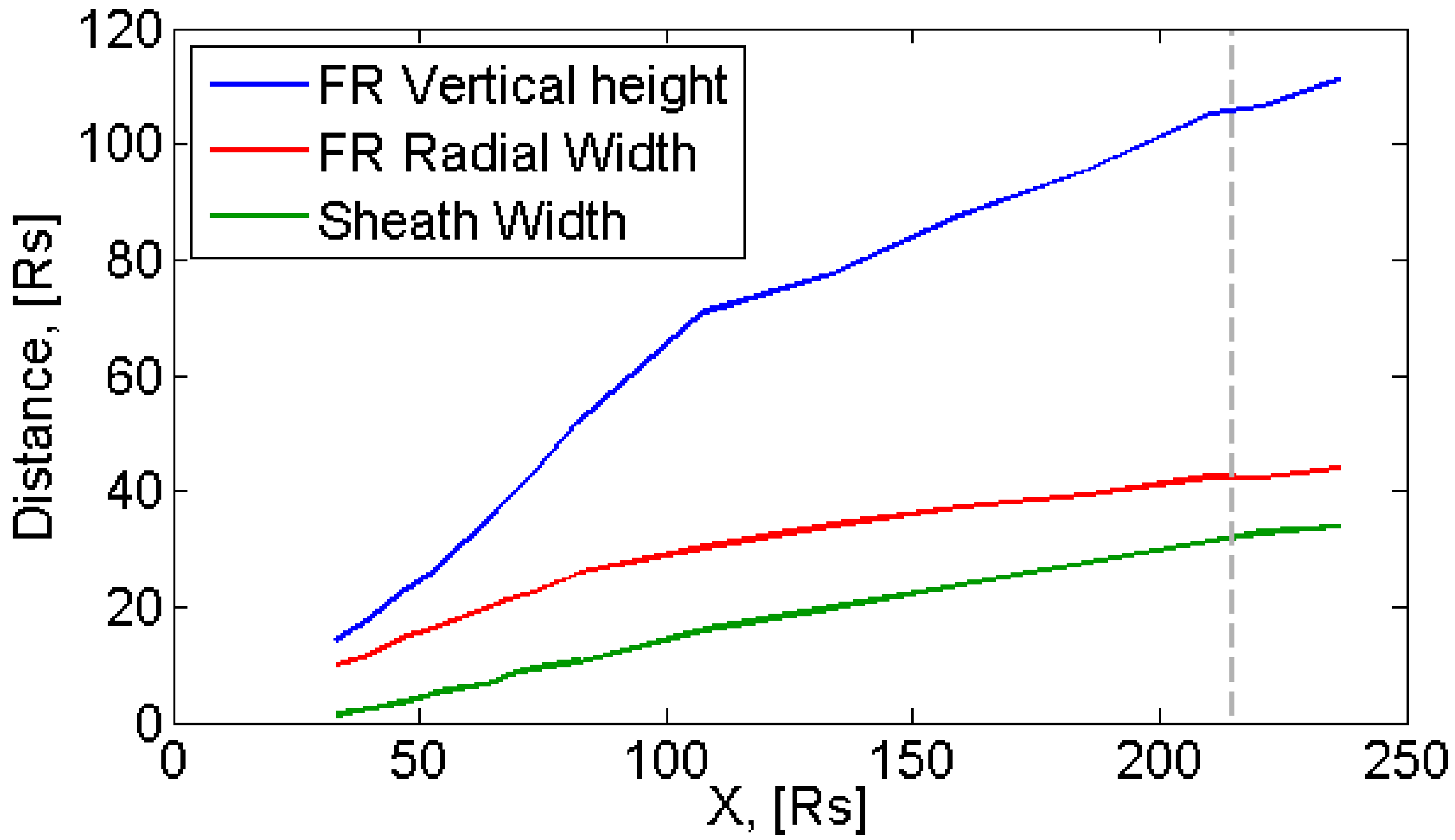

Figure 7. An investigation of Case 1. The growth of a FR by tracking the increase in the radial width (red), vertical height perpendicular to the radial (blue) and the shock standoff distance (green). The vertical grey dashed line plots the heliocentric distance of Earth.

An analytical investigation of the sheath width (shock standoff distance) normalized to the radius of curvature of the leading edge was carried out by (Siscoe & Odstrcil 2008) with a focus on high Mach number (M > 5) conditions. We approximate the radius of curvature to the vertical height of the flux rope in our study and display the results as a function of heliocentric distance in figure 8. Siscoe and Odstrcil investigated the behaviour of the two types of sheath regions for ICMEs and magnetospheres, labelled 'propagation sheath' and 'expansion sheath', respectively. They conclude that, depending on the fraction of the weighted average between them, the weighted average normalized standoff distance should lie between 0.07 and 0.2. Figure 8 shows that cases 1-3 behave similarly, while case 4 is slightly large. The large expansion and high Mach number during the early propagation phase generates a normalized distance of ~0.1. As the bulk flow of the flux rope decelerates, the Mach number drops; this behaviour along with the possibility that a larger weighting for the 'propagation sheath' should be included at the later stages of a CME's propagation might explain the growing normalized distance towards 0.3. These results are consistent with simulations carried out in the low corona below six solar radii (Loesch et al. 2011).

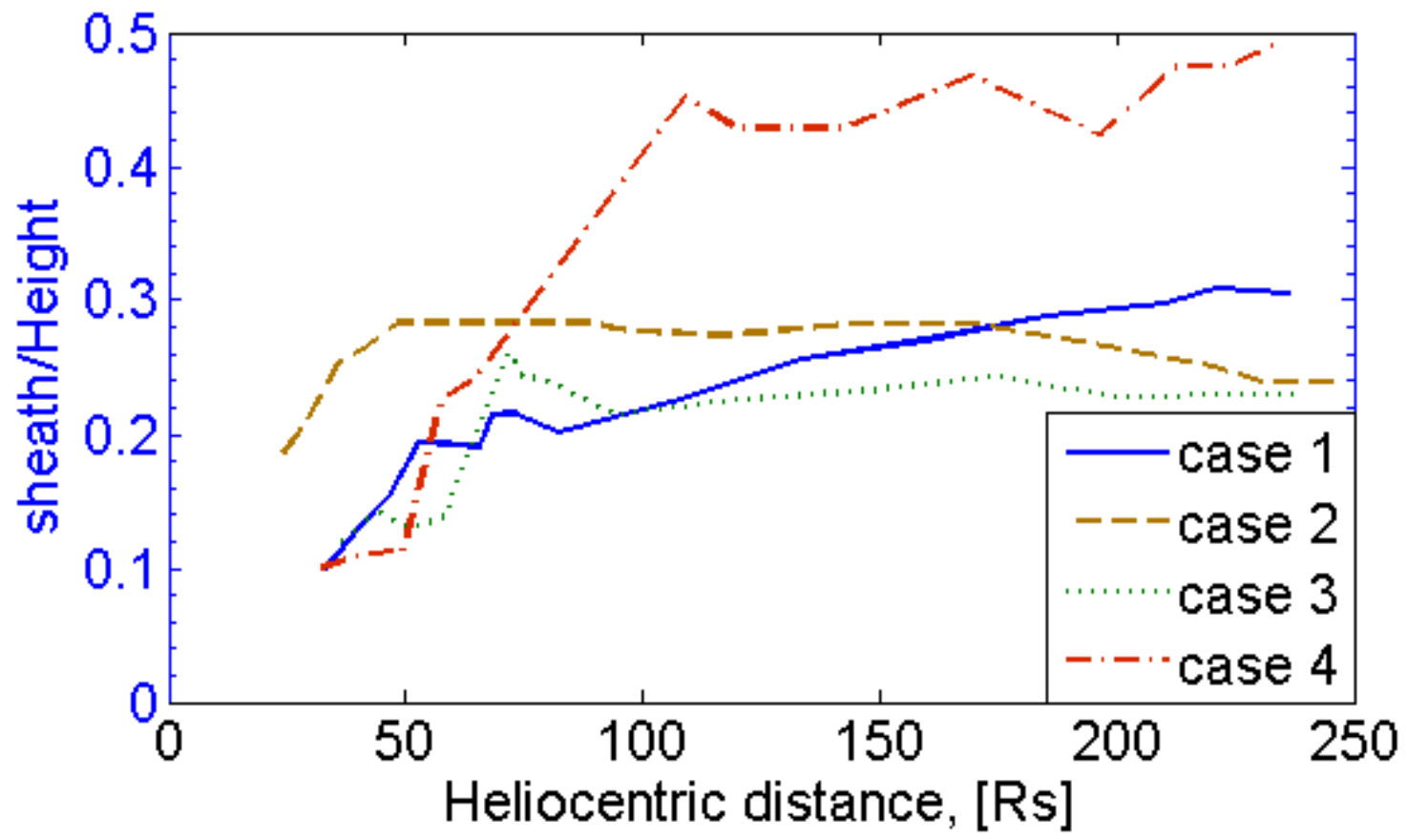

Figure8. An investigation of cases 1-4. The shock standoff distance is normalized to the vertical height of the flux rope (which is used as a proxy to the radius of curvature).

Figure 9 displays the ratio of plasma density across the shock front for the four simulations investigated in this study. The shock jump compression ratio (red dashed line) can be considered as a good proxy for the shock strength (see equation 3 and 4), and is shown to decrease in a non-linear way as the flux rope propagates into the heliosphere. However, we should note that the flux rope and the plasma parameters are not investigated until ~25Rs. This is because the flux rope is unrealistically initiated much further into the heliosphere than is observed and the sudden acceleration of the flux rope in the early phase would not be considered as typical propagation. The panels a – d represent the calculated parameters for case 1 –4 respectively. In this study we investigate the DR coefficient, $k_{dr}$, by re-arranging equation 1. This enables us to avoid the complications that arise from using different specific heat ratios or from using the mathematical derivations of Mach number from HD or MHD (equation 3 and 4). In addition, equation 1 is not explicitly related to the vertical height of the FR. As we wish to capture the behaviour of the shock and the appropriate jump conditions the specific heat ratio, $\gamma$, is set to 1.4. This is slightly below the 5/3 value of a thermodynamic monatomic ideal gas, therefore the theoretical maximum compression ratio is 6 and not 4 in our simulation. This is evident when inspecting the early stages of propagation for Case 4 (panel d), where the plasma compression ratio peaks to a little over 5.

The DR coefficient (blue solid line) in figure 9 shows a clear dependence on heliocentric distance. The popular value of 1.1 from equation 1, recently used to investigate in situ studies of CMEs (Savani, et al. 2011b) and remote observations (Gopalswamy, et al. 2012), is displayed as a horizontal dotted grey line. The lower bound appears to be consistent for all the simulations and remains approximately 0.8±0.1 at small heliocentric distances. This value supports equation 2 and the spherical geometry of a magnetosphere studied by Seiff (1962) from the Sun to a heliocentric distance of about 30Rs (~0.15AU). $k_{dr}$ appears to increase linearly with heliocentric distance as can be seen by the linear best fit curve displayed as a blue dot-dashed line. The linear approximation appears to improve as the initial speed of the injected CME increases. We hypothesize that the change in $k_{dr}$ is due to the change in the CME shape, which becomes more oblate as it propagates (indicated by the increase in aspect ratio). Panel d appears to contradict this hypothesis, by showing an aspect ratio that remains predominately constant at a little above 1. However, caution should be taken when considering panel d because the large injected speed of the CME (1200km/s) is a limitation on our simulation. The effect of this large speed on our simulation is to dramatically increase the radial expansion of the CME, which is perhaps not entirely realistic to observations even though evidence of many CMEs with small aspect ratios have been made (Lynnyk, et al. 2011; Savani, et al. 2011b). So the linear increase in $k_{dr}$ is more accurately linked, not to the aspect ratio, but to the size and the radius of curvature of the CME's leading edge.

Panel a, b, and c have been set to the same x/y axes for the ease of comparing $k_{dr}$ for the different simulations. Panel a and c show that an increase in speed which increases the density compression ratio across the shock and the aspect ratio of the CME has little effect on $k_{dr}$ ($k_{dr}$ is about 1.8 ±0.3 at Earth for both simulations). These two simulations also show that the coefficient remains predominately above the nominal 1.1 during the CME's propagation to 250Rs, achieving $k_{dr}$=1.1 at approximately 80Rs.

Case 2 (Panel b) displays the results for an initial flux rope with half the diameter, and hence half the angular width, of case 1 (panel a). Here we see a significant difference between the $k_{dr}$ values even though the aspect ratio of both CME's appear to be similar. This means the difference is due to the vertical extent of the CME, which supports our early suggestion that the radius of curvature (vertical extent) of the flux rope's leading edge is a significant variable that changes $k_{dr}$. Case 2 shows that for smaller CMEs with an angular width of ~22.6°, a $k_{dr}$ value of 1.1 appears to be adequate between the Sun and 215Rs as a first approximation because the value only varies between 0.8 and 1.25.

Finally, our results are in good agreement with density ratios derived by remote observations of CME-driven shocks at heights below 20Rs (Ontiveros & Vourlidas 2009). They obtained density compression ratios of >1.5 for distances above 12 Rs. If we combine the Ontiveros and Vourlidas (2009) measurements with our results, we conclude that the density compression ratio should increase up to a height of about 20-30Rs and then begin to decrease. This implies that CMEs tend to decelerate above 30Rs. Further remote measurements of density compression ratios, especially in the fields of view of heliospheric imagers (>40Rs), will be highly valuable in verifying our simulation results.

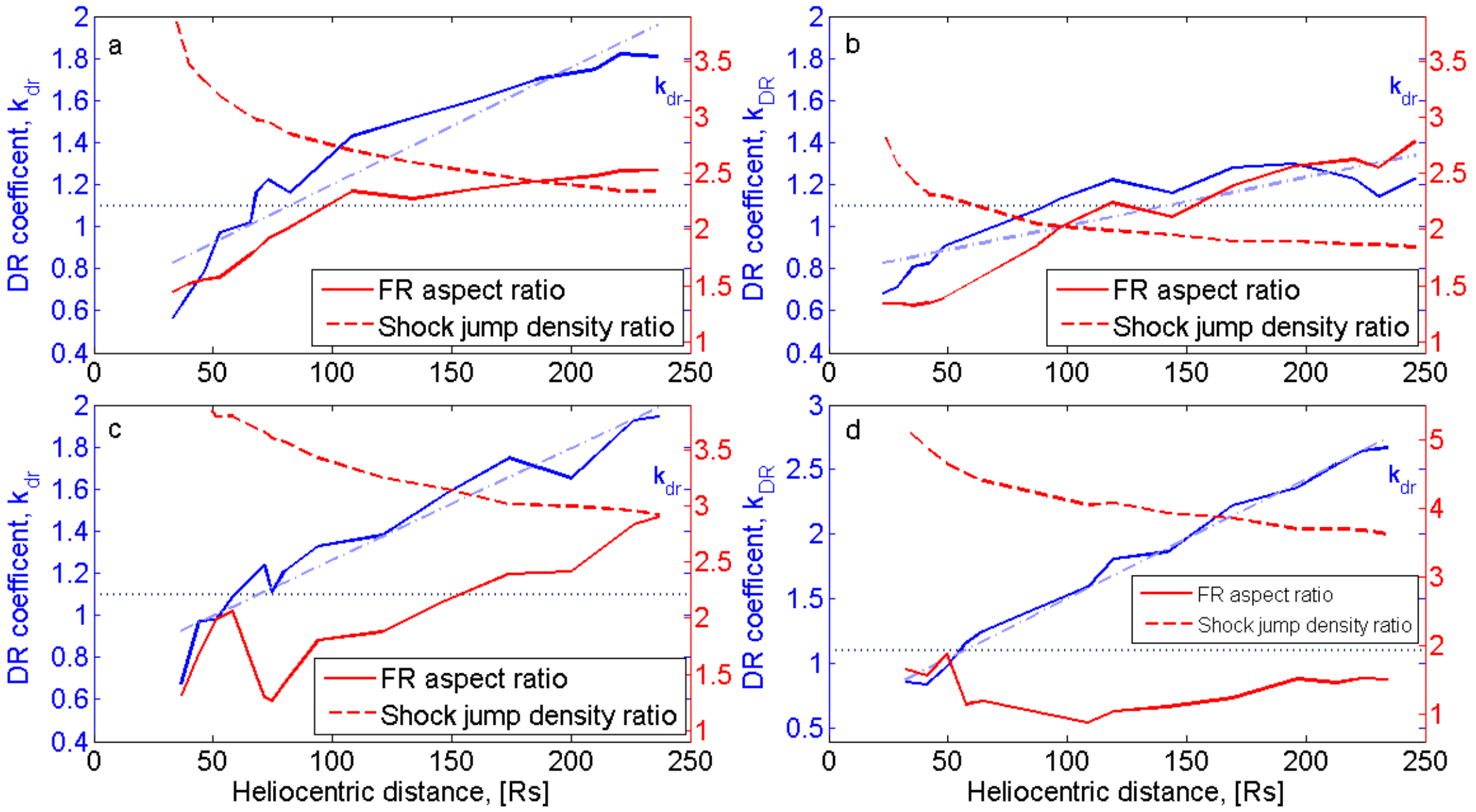

Figure 9. The aspect ratio (red solid line), plasma density compression ratio across the shock (red dashed line) and the DR coefficient (blue solid line) from equation 1 are displayed as a function of heliocentric distance, X. The blue and red curves relate to the left and right vertical y-axis, respectively. The linear best fit for $k_{dr}$ is displayed as a dot-dashed blue line, while the expect value is the grey dotted line at 1.1. Panels a, b, c and d correspond to case 1, 2, 3 and 4, respectively.

# 5. Conclusions

In this paper we investigate the behaviour of a CME and its associated shock standoff distance with the aid of 2.5D simulations between 10 and 300 Rs. By examining the density compression ratio coefficient, $k_{dr}$ (from equation 1), we investigate the appropriateness of comparing the geometry of Earth's magnetosheath to a sheath found downstream of a CME-associated shock. We have chosen to investigate this problem with 2.5D simulations rather than the 3D case for reasons of computational speed, our ability to simply repeat the problem for different initial parameters and because the scientific study concentrates only on the cross-sectional cut through the axis of a flux rope. However, we note a comparison between 2.5D and 3D simulations have been made (Jacobs, et al. 2007), and have concluded that the results are similar when their injected radial momenta are the same.

We conclude that a comparison between Earth Magnetosphere and a magnetic flux rope like structure, as expressed in equation 1, is appropriate as a first approximation, in particular for narrow CMEs. But also, that the geometry of a CME changes during its propagation and therefore the $k_{dr}$ value changes. It is more appropriate to use a $k_{dr}$ value of 0.8±0.1 for small heliocentric distances (i.e. for remote observations between 1 and 30Rs). This lower $k_{dr}$ value corresponds to an estimated spherical geometry of Earth's magnetosphere (equation 2, Seiff 1962). This geometry and a value of 0.8 was shown to be successful at estimating the magnetic field strength in the corona where the CME was observed and measured as a circular structure (Kim et al. 2012). As the CME propagates the cross section becomes more oblate and the $k_{dr}$ value increases linearly with heliocentric distance, such that $k_{dr}$= 1.1 is most appropriate at a heliocentric distance of about 80Rs. For terrestrial distances (215Rs) we estimate $k_{dr}$= 1.8 ±0.3, which also indicates that the CME structure is generally more oblate than Earth's magnetosphere.

In the four simulated case studies, we investigated the effect of initial speed and the size of the injected flux rope. We can see that the $k_{dr}$ value is significantly affected by the CME's heliocentric distance and by the size (radius of curvature) of the leading edge of the magnetic flux rope. The initial speed appears to have only a small effect, with the exception of case 4, which may reflect a limitation of our model, namely due to excessive radial expansion created from an instantaneous initial flux rope speed. However, as the initial speed of the flux rope was increased the uncertainty in the linear best fit curve for $k_{dr}$ values (as a function of distance) decreased. The magnetic field strength within the flux rope (between case 1 and 2) also had a less significant effect.

Throughout the analysis, we have concentrated on the height of the CME as a proxy for the radius of curvature of the leading edge in the plane of the cross section. Realistic CMEs are 3D objects that also possess a curvature of the leading edge that is perpendicular to the cross section. We have chosen to ignore this effect here. In essence, we have assumed that the radius of curvature in this dimension is significantly larger than the cross section curvature, which we can therefore assume to be linear. As our simulation is 2.5D, our propagating flux rope is in fact a 3D torus whose perpendicular curvature is the same as the heliocentric distance.

This work also concentrated on the radial width of a CME along the center of the flux rope. It is extremely unlikely that any spacecraft would travel exactly through the center, and therefore our analysis is idealised. Also for the simplicity of measuring the aspect ratio of the flux rope at specified frames, we did not measure the radial width of the CME as a time series. Due to a flux rope expanding while it passes over a fixed position, a difference of approximately 5% is expected between the radial widths of the two different methods (Owens, et al. 2006). Our method is therefore less appropriate for any comparison to in situ measurements but more relevant to the underlying physics for $k_{dr}$ values. Also recent observations (Nieves-Chinchilla, et al. 2012; Vourlidas et al. 2011) and simulations (Shiota, et al. 2010) have shown that CMEs may rotate as they propagate; this would also complicate any attempt to estimate the aspect ratio from in situ measurements.

Any future investigation should try to mitigate the limitations of our work by focusing on the flux rope structure at smaller heliocentric distances (<30Rs), so that more detailed comparisons can be made for remotely observed CMEs (Gopalswamy, et al. 2012). Also, an improvement on our instantaneous injection of a fast flux rope can be made by introducing a flux rope with a more realistic acceleration over a short distance. This will allow future work to study larger maximum speeds and therefore investigate extreme events that are more likely to be significantly geo-disruptive.

**Acknowledgments.** NPS was supported by the NASA Living With a Star Jack Eddy Postdoctoral Fellowship Program, administered by the UCAR Visiting Scientist Programs and hosted by the Naval Research Laboratory. NPS and AV were supported by NASA and the office of Naval Research. The numerical calculations were made using the supercomputing cluster at the Solar Terrestrial Environment laboratory, Nagoya University.